\documentclass[%
 aip,
 amsmath,amssymb,
 reprint,%
]{revtex4-1}
\usepackage{graphicx}
\usepackage{dcolumn}
\usepackage{bm}
\usepackage[utf8]{inputenc}
\usepackage[T1]{fontenc}
\usepackage{mathptmx}
\usepackage{etoolbox}
\usepackage[T1]{fontenc}
\usepackage{tikz}
\usepackage{color}
\usepackage{amsmath}
\usepackage{amssymb}
\usepackage{epsfig}
\usepackage{xcolor}
\definecolor{mynicegreen}{RGB}{102,182,102}
\makeatletter
\def\@email#1#2{%
 \endgroup
 \patchcmd{\titleblock@produce}
  {\frontmatter@RRAPformat}
  {\frontmatter@RRAPformat{\produce@RRAP{*#1\href{mailto:#2}{#2}}}\frontmatter@RRAPformat}
  {}{}
}%

\begin{document}

\title{Effect of clustering on the orientational properties of a fluid of hard right isosceles triangles}

\author{Yuri Mart\'{\i}nez-Rat\'on}
\email{yuri@math.uc3m.es}
\affiliation{
Grupo Interdisciplinar de Sistemas Complejos (GISC), Departamento
de Matem\'aticas, Escuela Polit\'ecnica Superior, Universidad Carlos III de Madrid,
Avenida de la Universidad 30, E-28911, Legan\'es, Madrid, Spain}

\author{Enrique Velasco}
\email{enrique.velasco@uam.es}
\affiliation{Departamento de F\'{\i}sica Te\'orica de la Materia Condensada,
Instituto de F\'{\i}sica de la Materia Condensada (IFIMAC) and Instituto de Ciencia de
Materiales Nicol\'as Cabrera,
Universidad Aut\'onoma de Madrid,
E-28049, Madrid, Spain}

\date{\today}

\begin{abstract}
	Recent studies have shown the fluid of hard right triangles to possess fourfold and quasi-eightfold (octatic) orientational symmetries. However, the standard density-functional theory for two-dimensional anisotropic fluids, based on two-body correlations, and an extension to incorporate three-body correlations, fail to describe these symmetries. To explain the origin of octatic symmetry we postulate strong particle clustering as a crucial ingredient. We use Scaled Particle Theory to analyze four  binary mixtures of hard right triangles and squares, three of them being extreme models for a one-component fluid, where right triangles can exist as monomeric entities together with triangular dimers, square dimers or square tetramers. Phase diagrams  exhibit a rich phenomenology, with demixing and three-phase coexistences. More important, under some circumstances the orientational distribution function of triangles has equally high peaks at relative particle angles $0,\ \pi/2,$ and $\pi$, signalling fourfold, tetratic order, but also secondary peaks located at $\pi/4$ and $3\pi/4$, a feature of eightfold, octatic order. Also, we extend the binary mixture model to a quaternary mixture consisting of four types of clusters: monomers, triangular and square dimers, and square tetramers. This mixture is analyzed using Scaled Particle Theory under the restriction of fixed cluster fractions. Apart from the obvious tetratic phase promoted by tetramers, we found that, for certain cluster compositions, the total orientational distribution function of monomers can exhibit quasi-eightfold (octatic) symmetry. The study gives evidence on the importance of clustering to explain the peculiar orientational properties of liquid-crystal phases in some two dimensional fluids.
\end{abstract}

\maketitle

\section{Introduction}
\label{introduction}

The experimental and theoretical study of two-dimensional (2D) fluids of hard anisotropic particles has
enjoyed an upsurge in recent years, mainly motivated by the
development of novel experimental techniques such as lithographic particle fabrication
\cite{Zhao1,Zhao2,Zhao3,Zhao4}. Using these techniques, micro-prisms of any cross-sectional shape can 
be fabricated, and suspensions of these particles can be adsorbed on surfaces, giving rise to effectively 
two-dimensional fluids of diffusing Brownian particles \cite{Zhao1,Zhao2,Zhao3,Zhao4}. 
The fluid phase behavior can be explored by varying particle volume fraction, and 
in many cases a plethora of exotic liquid-crystal and crystalline phases results. These phases possess
symmetries that strongly depend on particle shape. Some of these phases were predicted theoretically
and later confirmed by Monte Carlo (MC) simulations, and different theoretical models have been developed to 
explain the rich phase behavior of these 2D hard-core fluids and its particle shape dependence
\cite{Schlaken,Frenkel,Donev,Dijkstra,MR1,MR2,MR3,MR4,Dani,Escobedo,Glotzer1,Cinacchi}. 

Of particular interest are the triatic (TR) and tetratic (T) phases found in fluids of hard equilateral 
triangles \cite{Zhao4,Dijkstra,MR3} and squares (and also in rectangles of small aspect ratios) \cite{MR2,Escobedo}, 
with particle axes pointing along six or four equivalent directors, respectively. The T 
phase can be viewed as the 2D analog of the biaxial nematic phase, recently found to be stable in colloidal 
suspensions of board-like particles \cite{Vroege,Vanakaras} and whose
stability can be enhanced by size polydispersity \cite{Roij,Patti,Patti2}. 

Vertically vibrated granular monolayers are being studied as experimental models of real 2D fluids in thermal
equilibrium. Under specific experimental conditions, monolayers of squares 
\cite{experiments} and cylinders \cite{Dani2,MR4b,MR5} exhibit the presence of T and also 
smectic liquid-crystal textures in the steady-state configurations. The excitation of topological defects 
in the orientational and positional director fields of these fluidized granular monolayers, when confined into 
circular cavities, have been observed and studied \cite{MR5}. These results point to the 
importance of hard core entropic interactions in the stability of these dissipative textures, which turn out to be
very similar to those obtained in equilibrium experiments on monolayers of colloidal spherocylinders confined 
in cavities of different shapes \cite{Wittmann1,Wittmann2}. This connection opens up the possibility that vibrated granular 
monolayers may be considered as valid experimental models to probe the interplay between symmetry and order in 2D fluids.

The most successful theoretical tool used in the study of liquid-crystal phase behavior of 
hard-body fluids is Density Functional Theory (DFT) \cite{hard-body}. The main advantage of DFT is that
it allows to obtain, via functional minimization, the equilibrium angular distribution function $h(\phi)$
of a 2D oriented fluid, i.e. the probability density of particle axes to orient at an angle
$\phi$ with respect to one of the equivalent directors. As shown in Ref. \onlinecite{MR3}, the Scaled-Particle Theory (SPT)
version of DFT (which includes only two-body correlations) 
predicts that the uniaxial nematic (N) phase is the only stable orientation phase of a fluid
of hard right triangles, i.e. no exotic liquid-crystal phases do exist in this fluid. A bifurcation analysis,
corroborated by rigorous functional minimization close to the bifurcation, confirmed this result, and
the incorporation of three-particle correlations did not modify this scenario \cite{nosotros}.
By contrast, MC simulations of the same fluid showed the presence of the T phase 
(obtained by expanding a perfect T-crystal), along with an additional exotic oriented fluid phase 
with eightfold symmetry, the octatic (O) phase, obtained by compressing the isotropic (I) fluid. 
Even though all evidence suggests T to be the true stable phase, the fluid is prone to developing strong O 
correlations as density is increased from the I fluid. Note that the T and O phases are highly-symmetric
phases having fourfold and eightfold symmetries, i.e. their angular distribution functions have the property 
$h(\phi)=h(\phi+2\pi/n)$, with $n=4$ and 8, respectively \cite{Dijkstra,nosotros}. 
The failure of the standard Onsager theory and its variations, all based on two-body correlations, to
predict the phase behaviour of hard-particle models, at least at a qualitative level, is quite
unusual in the history of liquid crystals. In Ref. \onlinecite{nosotros} we advanced a reason why the standard 
two-body theory, and also the three-body-extended theory, cannot predict the highly-symmetric T and O phases,
namely the crucial contribution of fourth-, and probably even higher-order, correlations in this system. 

Given the difficulty of improving the standard theories by incorporating such high-order correlations, in this work
we explore different ideas in an effort to understand the problem from different perspectives.
On the one hand we focus on a system that should promote 
orientational correlations with O symmetry: a binary mixture of hard right triangles and squares. 
The excluded area between these two particles shows 
local minima at relative angles $\phi=\pi/4$ and $3\pi/4$, which could
drive a stable O phase. As will be seen, this property of the excluded area is not enough to 
promote bulk O ordering, and the N and T phases are the only oriented phases that get stabilized
in the phase diagrams for the four different mixtures analyzed. Despite this, we found that the  
orientational distribution function of triangles, for certain values of mixture composition, 
exhibits small secondary peaks located at the relative angles above.

On the other hand, we study the
effect of particle clustering in the orientational properties of hard right triangles. Clustering
is an extreme consequence of high-order particle correlations and could be a complementary point
of view to extract useful information on the fluid behavior. We formulate a theory for clustering
with the help of a toy model for particle self-assembling where monomers are just "free" right 
triangles. These monomers are assumed to self-assemble into different triangular and square-shaped clusters, 
the latter coming in two varieties: dimers and tetramers. An effectively quaternary mixture
results from these considerations, which is analyzed using the SPT version of DFT.
We numerically minimize the functional for particular compositions and, from the equilibrium angular 
distribution functions of the four species, a monomer distribution function $h_{\rm m}(\phi)$ can be 
predicted. It is then shown that, for certain cluster compositions, this function 
has quasi-eightfold symmetry, i.e. four peaks of similar heights at $\phi=k\pi/4$ ($k=0,\cdots,3$) 
in the interval $[0,\pi]$. This result demonstrates the relevance of clustering to explain the presence 
of O orientational symmetry in a fluid of right triangles. 

Aside from the theoretical calculations, we have also performed MC simulations of the real fluid
of right triangles to confirm the presence of clustering. It has been pointed 
out \cite{Glotzer2} that entropic interactions between anisotropic particles in dense fluids can 
in some sense be regarded as chemical bonds, that in turn may promote particle self-assembling. 
In our simulations we define a criterion to identify different clusters: triangular, square and 
rhomboidal dimers, and also square tetramers. Cluster fractions are analyzed as a function of 
packing fraction in MC compression runs starting from the I fluid, and also from expansion runs 
from the T crystalline phase. We show that the T phase is enriched in square dimers and tetramers,
with a small proportion of the remaining clusters. By contrast, all clusters have similar fractions
in the O phase. 

Instead of the fixed cluster compositions assumed in our toy model, a more
sophisticated model to describe strong clustering effects in hard particle fluids 
should certainly predict cluster compositions at equilibrium in a consistent fashion. Chemical equilibrium between 
different clusters, a mass conservation law, and a larger variety of clusters
(such as clusters with rhomboidal shape), along with effective internal energies of clusters
are important ingredients that the new model should take into account. This line of research 
we leave for future developments.

The article is organized as follows.  In Sec. \ref{mixtures} we study four different 
binary mixtures of hard squares and hard right triangles and calculate 
their phase diagrams and the orientational properties of the different species. 
The effect of clustering on the stability of the liquid-crystal phase with eightfold 
symmetry is analyzed in Sec. \ref{clustering}. MC simulations and results for cluster fractions are shown
in Section \ref{MC}. Finally some conclusions are drawn in Sec. \ref{conclusions}.

	\section{Binary mixtures of right triangles and squares}
	\label{mixtures}

	\begin{figure}
		\hspace*{0.8cm}
		\epsfig{file=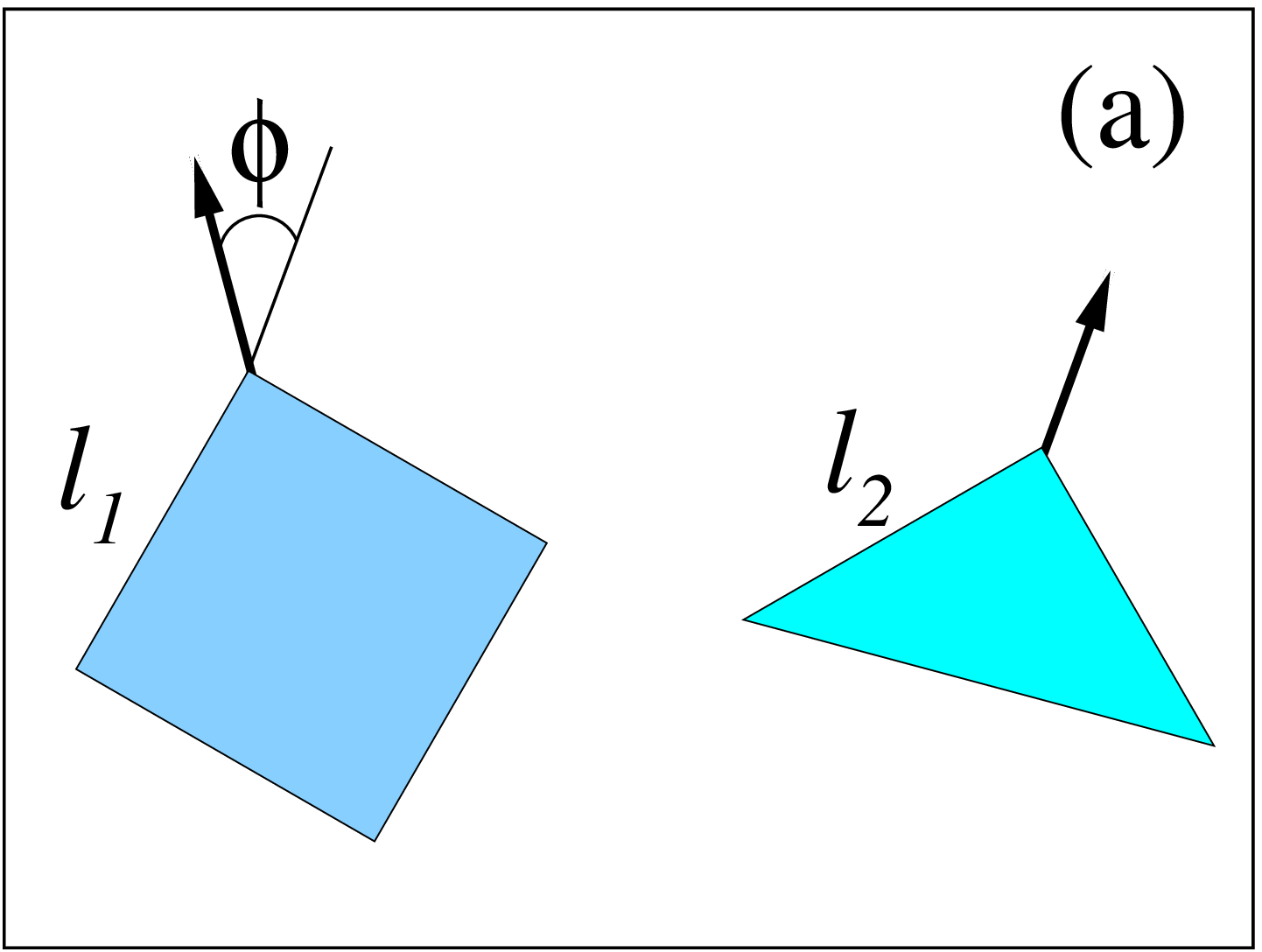,width=1.7in}
		\epsfig{file=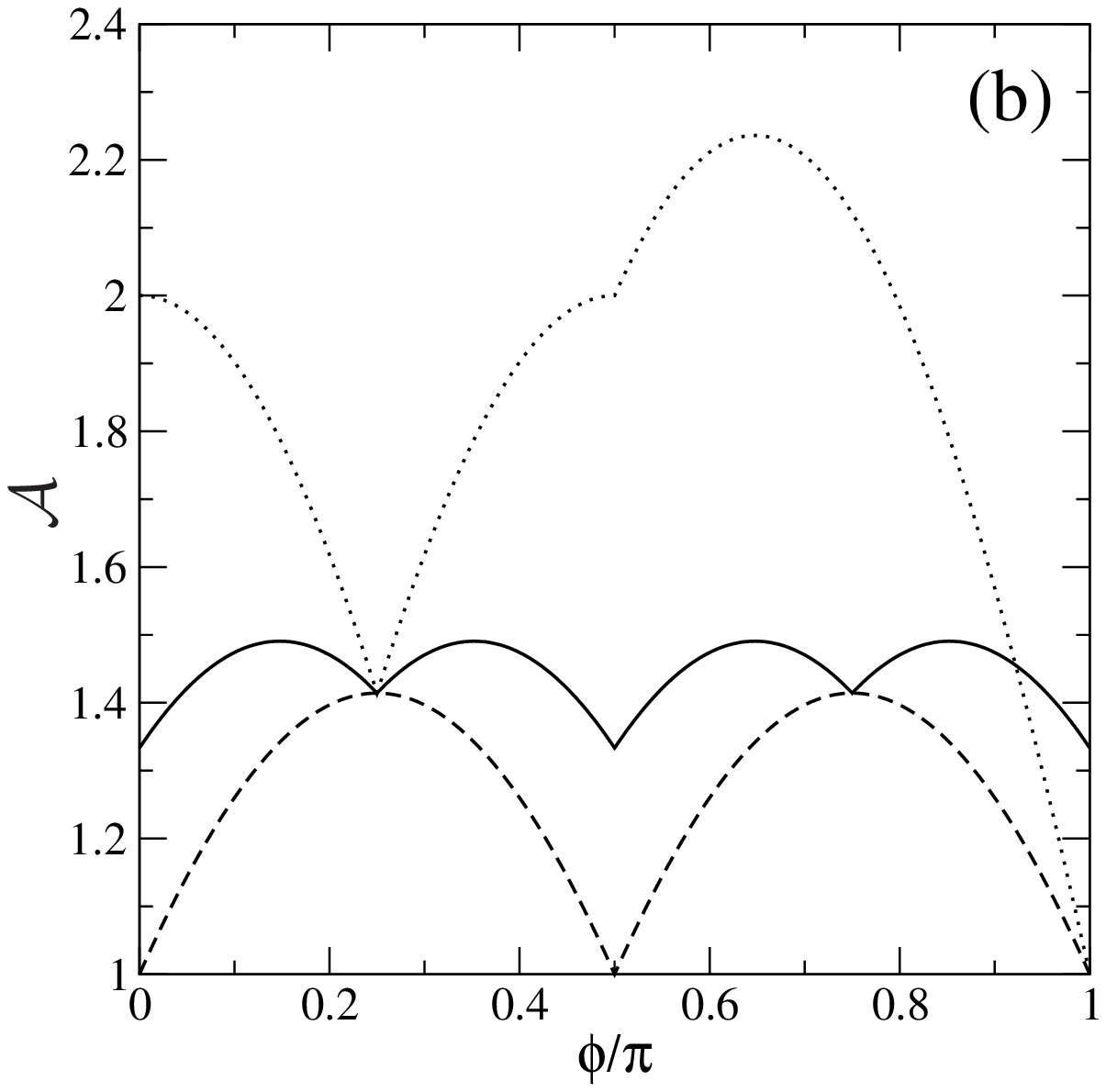,width=3.in}
		\caption{(a) Sketch of square (species 1) and right-angled triangle (species 2) 
		with a relative angle $\phi$ between their main axes. The equally-sized lengths 
		of particles $l_1$ and $l_2$ are respectively shown. 
		(b) The scaled excluded area ${\cal A}(\phi)\equiv
		\left({\cal K}_{ij}^{(2)}(\phi)-a_i-a_j\right)/(a_i+a_j)$ between a square and a right 
		triangle (solid curve), between two squares (dashed curve), and between two right 
		triangles (dotted curve). In this example both
		particles have the same equally-sized lengths, $l_1=l_2$,
		so that the area of a square is twice that of a triangle.}
		\label{fig1}
	\end{figure}

	In this section we study the phase behavior and orientational 
	properties of binary mixtures of squares (species 1) and right triangles (species 2). 
	The motivation is that the cross excluded area ${\cal K}_{12}^{(2)}(\phi)$ (apart from 
	being symmetric with respect to $\phi=\pi/2$) exhibits four local minima in the interval 
	$[0,\pi)$, located at relative angles $\phi=k\pi/4$ ($k=0,\cdots,3$); see Fig. \ref{fig1} where
        the relative angle $\phi$ is defined and the excluded area shown for the particular case 
	of squares and triangles of the same (equally-sized) lengths, $l_1=l_2$ (solid curve in the figure).
	We can see however that the gain in scaled excluded area for these relative angles
	is rather modest. For comparison the same figure shows the excluded area between like species.
	In these cases the gain in excluded area at the local minima is much higher. 
	Despite this, the presence of four local minima in ${\cal K}_{12}^{(2)}(\phi)$ could in turn promote 
	the stability of the O phase in the binary mixture. Even though the standard DFT 
	does not predict the stability of the O phase in one-component fluids \cite{nosotros}, 
	the mixing of right triangles with other particles such that the cross excluded area
	presents eight local minima in the interval $[0,2\pi)$ could generate an O phase.
	One example of such a particle is the square. We remind the reader that simulations, 
	DFT and experimental studies on vibrated monolayer of hard squares 
	\cite{Frenkel,MR1,experiments} have shown that the one-component fluid 
	exhibits I-T and T-crystal second-order phase transitions.

	To analyze this mixture we used a DFT based on the SPT-second virial 
	theory, generalized to binary mixtures. The proposed expression for the excess 
	free-energy per particle in reduced thermal units is 
	\begin{eqnarray}
		&&\varphi_{\rm exc}[\{h_i\}]=-\log(1-\eta)+\frac{\rho}{2(1-\eta)}
		\nonumber\\
		&&\times \sum_{i,j}x_ix_j\left({\cal K}^{(2)}_{ij,0}-a_i-a_j+\frac{1}{2}
		\sum_{n\geq 1}^N 
		{\cal K}^{(2)}_{ij,n} h_n^{(i)}h_n^{(j)}\right).
		\label{proposed}
	\end{eqnarray}
	Here a truncated Fourier expansion of the orientational distribution functions,
	\begin{eqnarray}
		h_i(\phi)=\frac{1}{2\pi}\left(1+\sum_{n\geq 1}^N h_n^{(i)}\cos(2n\phi)\right),
	\end{eqnarray}
	is used to calculate the double angular average with respect to $h_i(\phi)$ and 
	$h_j(\phi')$ in the SPT expression \cite{nosotros0} 
	${\cal K}_{ij}^{(2)}(\phi-\phi')-a_i-a_j$,
	giving the term between parenthesis in Eqn. (\ref{proposed}). The total 
	packing fraction is defined as $\displaystyle\eta=\rho\left<a\right>$,
	i.e. the product of the total number density $\rho$ and the average area
	$\displaystyle\left<a\right>\equiv\sum_i x_i a_i$, given by a sum over species of the products 
	of molar fractions $x_i$ and particle areas $a_i=l_i^2/i$ ($i=1,2$). Here 
	$l_i$ is the equally-sized side-length of species $i$ (see Fig. \ref{fig1}).
The coefficients ${\cal K}^{(2)}_{ij,n}$ can be computed analytically from the expressions 
\begin{eqnarray}
	&&{\cal K}^{(2)}_{ij,n}-(a_i+a_j)\delta_{n0}\nonumber\\
	&&=
	-\frac{4l_il_j\left[1+\delta_{i2}\delta_{j2}+
	(-1)^n+(\delta_{i2}+\delta_{j2})\sqrt{2}\cos\left(\frac{n\pi}{2}\right)\right]}
	{2^{\delta_{i2}+\delta_{j2}}(4n^2-1)\pi}.\nonumber\\
	&&
	\label{coefficients}
\end{eqnarray}
The ideal part of the free-energy for the mixture is 
\begin{eqnarray}
	\varphi_{\rm id}(\{h_i\})=\log \eta -1 +\sum_i 
	x_i \left[\log x_i +\int_0^{2\pi}h_i(\phi)\log h_i(\phi)\right].
	\nonumber\\
	\label{lo_ideal}
\end{eqnarray}
The total free-energy per particle is 
$\varphi(\{h_i\})=\varphi_{\rm id}(\{h_i\})+\varphi_{\rm exc}(\{h_i\})$, 
and the Gibbs free-energy functional per particle $g$ can be obtained from
\begin{eqnarray}
	g(\{h_i\})=\varphi(\{h_i\})+\frac{\beta p}{\rho}.
\end{eqnarray}
The latter expression is very useful for the calculation of the phase diagrams of binary mixtures, in particular 
when the fluid demixes into two coexisting phases. The procedure is: (i) fix the pressure 
to some constant value $p (x,\rho)=\rho^2\partial \varphi/\partial \rho=p_0$; (ii) using 
this constraint, calculate the total density $\rho(x;p_0)$ as a function of the molar fraction 
of squares $x\equiv x_1$, and insert back into the Gibbs free-energy to obtain 
the function  $g(x,p_0)$. Note that in the above procedure all Fourier amplitudes $\{h_n^{(i)}\}$ 
have to be obtained through the equilibrium condition $\partial \varphi/\partial h_n^{(i)}=0$.  
From the double-tangent
construction of the function $g(x,p_0)$, which guarantees the equality of chemical potentials 
of the species at the coexisting phases, we find the coexisting values $x^{(a)}$ and 
$x^{(b)}$, and from these the coexisting densities $\rho^{(a)}$ and $\rho^{(b)}$. 
Changing the pressure $p_0$ and repeating the above procedure we can construct that 
part of the phase diagram in the pressure-composition plane where demixing is present.   

In the case of second order phase transitions a bifurcation analysis is required (see \cite{MR3}
for the case of mixtures of triangles using the SPT formalism). The packing fraction at bifurcation
(spinodal curves)
turns out to be a simple generalization of the corresponding expression for the one-component fluid 
\cite{nosotros}:
\begin{eqnarray}
	\eta_n=\frac{1}{
		1-\sum_i x_i {\cal K}^{(2)}_{ii,n}/\langle a\rangle}.
	\label{spin2}
\end{eqnarray}
The orientational order of squares and triangles is measured using the set of order parameters 
\begin{eqnarray}
	Q_{2n}^{(i)}=\int_0^{2\pi}d\phi h_i(\phi)\cos(2n\phi)=\frac{h_n^{(i)}}{2}.
\end{eqnarray}
These parameters account for N ($n=1$), T ($n=2$), TR ($n=3$) and O ($n=4$) ordering.

\begin{figure}
	\epsfig{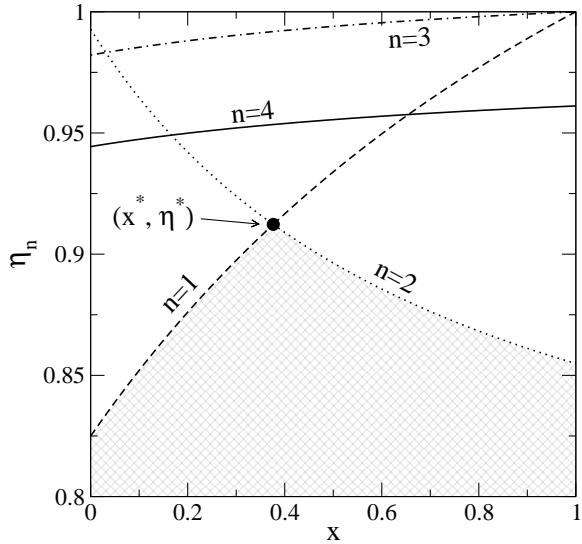}
	\caption{Bifurcation curves $\eta_n(x)$ for $n=1$ (I-N, dashed line), 
	$n=2$ (I-T, dotted line), $n=3$ (I-TR, dot-dashed line), and $n=4$ (I-O, solid line) 
	corresponding to squares and triangles of equal lengths,
	$l_1=l_2$. $x$ is the fraction of squares. The solid circle is the intersection point $(x^*,\eta^*)$
	of the I-N and I-T bifurcation curves. The patterned area indicates the
	region where the I phase is stable against orientational order
	}
	\label{fig2}
\end{figure}

\subsection{Bifurcation curves}

The spinodal curves $\eta_n(x)$ for $n=1,\cdots,4$ from Eqn. (\ref{spin2}) are 
plotted in Fig. \ref{fig2} for a binary mixture of particles with $l_1=l_2$. 
The I-N ($n=1$) and I-T ($n=2$) curves departing from $x=0$ (one-component triangle fluid) 
and $x=1$ (one-component square fluid) are monotonically increasing functions of $x_i$ 
($i=1$ for I-N and 2 for I-T respectively), and intersect at 
$x^*\simeq 0.376$. This in turn means that mixing stabilizes the I phase, which 
can be easily explained by the different (two- vs. four-fold) symmetries of the liquid-crystal 
phases of hard-triangle and hard-square fluids, respectively. The shaded area
in the figure indicates the region of I-phase stability against orientational order.
As we will shortly see, the point $(x^*,\eta^*)$ is always located inside the demixing gap. 
It is interesting to note from Fig. \ref{fig2} 
that the effect of mixing leads to the lowering of the packing-fraction difference 
between the I-O and I-N or I-T bifurcations. 
This indicates that the O phase of the mixture becomes "less" 
unstable with respect to the N or T phases. However this mixing effect  is not sufficient 
to stabilize it. 

\begin{figure*}
	\epsfig{file=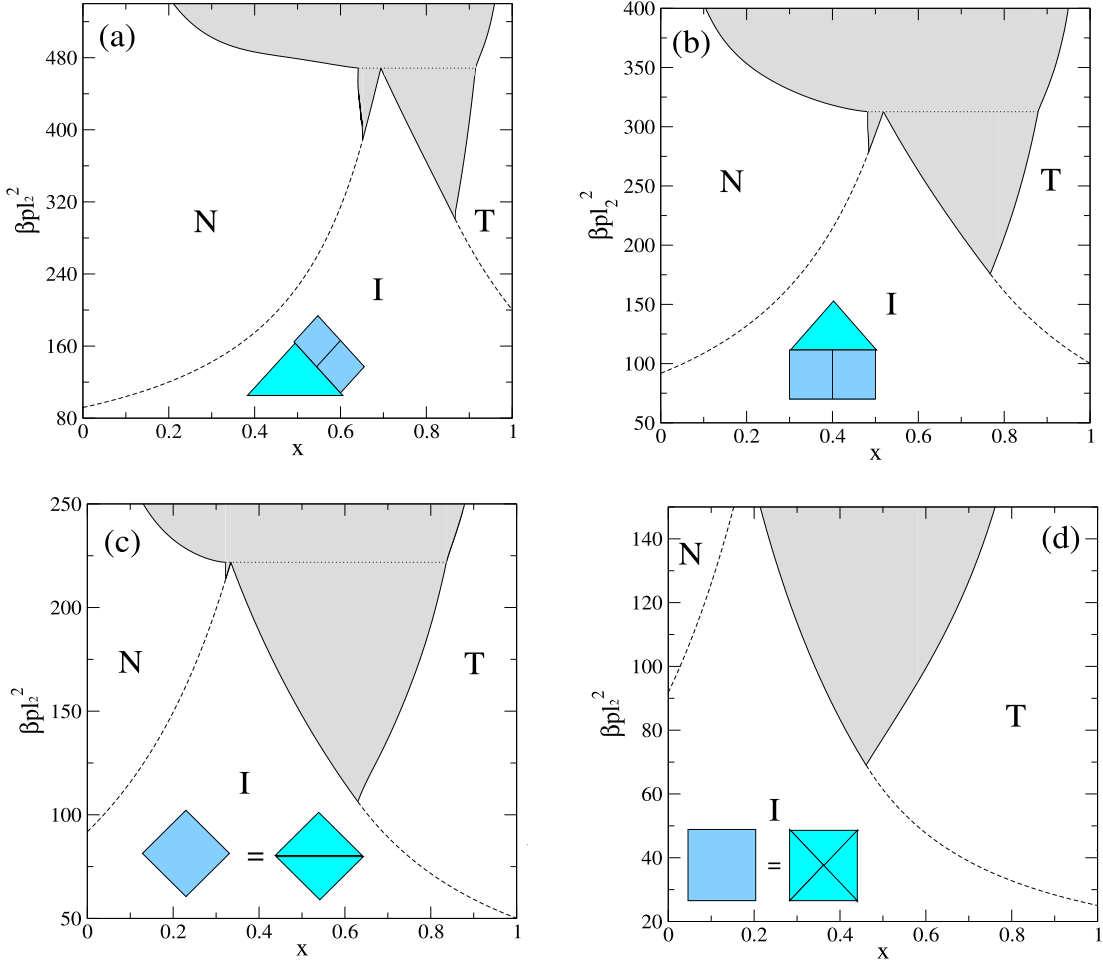,width=6.in}
	\caption{Reduced pressure-composition phase diagram of binary mixtures of hard squares 
	and triangles with (a) $(\kappa_l,\kappa_a)=(1/2,1/2)$, (b) $(1/\sqrt{2},1)$,
	(c) $(1,2)$, and (d) $(\sqrt{2},4)$. Coexistence binodals of first-order or demixing 
	transition are shown by solid curves, while dashed curves
	represent second-order I-(N,T) transitions. The regions 
	of stability of I, N and T phases are correspondingly labeled. The dotted horizontal line indicates
	the triple I-N-T coexistence. Particles are drawn in each case, showing the corresponding ratios 
	between particle areas.}
	\label{fig4}
\end{figure*}

\subsection{Phase diagrams}

Phase diagrams have been calculated for four different binary mixtures, Figs. \ref{fig4} (a)-(d). 
Defining the length ratio $\kappa_l\equiv l_1/l_2$, the four mixtures have
$\kappa_l=1/2$, $1/\sqrt{2}$, $1$, and $\sqrt{2}$. The area ratio $\kappa_a\equiv a_1/a_2$
for the mixtures is $1/2$, $1$, $2$, and 4, respectively. 
I-N and I-T second-order transition curves depart from the $x=0$ and $x=1$ axes, respectively.
Both curves end in corresponding tricritical points. For pressures above these tricritical
points the corresponding transitions become of first order. In the case of the I-T curve
the transition corresponds to strong demixing,
with strong fractionation of the two species.
Both phase transitions are bounded above by a triple I-N-T coexistence (dotted horizontal
lines in panels (a)-(c) of Fig. \ref{fig4}), and for higher pressures demixing takes place between a N phase, 
rich in triangles, and a T phase, rich in squares. It is interesting to note that the lowest
tricritical point is always that of the I-T spinodal curve. This is the effect of the
large decrement in total averaged excluded area, 
$\displaystyle\sum_{i,j}x_ix_j\langle\langle {\cal K}_{ij}^{(2)}\rangle\rangle_{h}$, 
when T (instead of N) orientational ordering is induced by squares.
It is clear from Fig. \ref{fig4}(b) that mixing of species with approximately the same areas 
(values of $\kappa_a$ in the neighborhood of unity) also exhibit strong demixing.
This can be explained by two facts, which we elaborate in the following.
\begin{itemize}
	\item[(i)] The different (two- vs. four-fold) symmetries of the 
N and T phases of the hard-triangle and hard-square fluids, respectively. 
Triangles oriented into a T configuration generate a high free-energy cost due to the 
particular form of the triangle-triangle excluded area. From Fig. \ref{fig1} we can see how 
the equipartition of particle orientations into the discrete set 
of angles $\{0,\pi/2,\pi\}$ (perfect T ordering) 
generates an averaged triangle-triangle scaled excluded area 
\begin{eqnarray}
	&&\langle\langle{\cal A}(\phi)\rangle\rangle\equiv
\frac{1}{2a}\langle\langle {\cal K}_{22}^{(2)}(\phi)\rangle\rangle_h-1\nonumber\\
	&&=\frac{1}{4}\left[{\cal A}(0)+2{\cal A}(\pi/2)
+{\cal A}(\pi)\right]=\frac{7}{4},
\end{eqnarray}
larger than that corresponding to equipartition into the angles $\{0,\pi\}$ 
(perfect N ordering), equal to 
$\displaystyle{\frac{1}{2}\left[{\cal A}(0)+{\cal A}(\pi)\right]=\frac{3}{2}}$. 
Also a fluid of hard squares cannot exhibit a N phase due to the symmetry of the particles,
which give an excluded area invariant under $\pi/2$-rotations. 
Therefore, at high enough pressures, phase-separation into two phases, each having the 
orientational order promoted by the most populated species, guarantees a much lower 
free-energy at equilibrium. 

\item [(ii)] The decrease in excluded area promoted by orientational order is less for the
triangle-square pairs than for triangle-triangle or square-square pairs
(Fig. \ref{fig1}), which obviously favors the demixed state.
\end{itemize}
A final comment on the phase diagrams is that the coexistence region of the first-order I-N
transition shrinks dramatically with the ratio $\kappa_l$, eventually disappearing
for $\kappa_l=\sqrt{2}$ (see panel (d) of Fig. \ref{fig4}).  
This is the 
most likely scenario, something that cannot be settled with total certainty as
the minimization in Fourier space cannot be achieved successfully for pressure 
values close to the intersection between the I-N second order transition curve 
and the N binodal of the N-T demixing. Our numerical minimization scheme does not give reliable 
results at these pressure values, even for a number of Fourier coefficients
$h_n^{(i)}$ equal to 100, due to the incorrect numerical
representation of $h_i(\phi)$.  
A pressure difference, measured from the I-T tricritical point, of
$\beta \Delta p l_1^2\simeq 160$, is the highest pressure for which 
we could perform accurate numerical minimizations.
If this scenario were correct the 
second-order I-N transition would end as a critical end-point at the N binodal of the N-T 
demixing transition.

\begin{figure}
	\epsfig{file=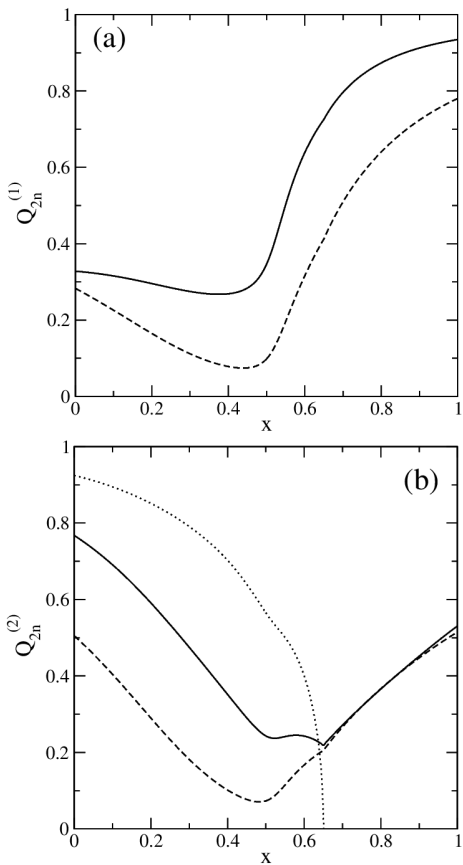,width=2.5in}
	\caption{Order parameters $Q_2^{(i)}$ (dotted), $Q_4^{(i)}$ (solid), and $Q_8^{(i)}$ (dashed) of 
	(a) squares ($i=1$), and (b) triangles ($i=2$), as a function of composition $x$. 
	The mixture asymmetry is $\kappa_l=1/\sqrt{2}$ ($\kappa_a=1$) 
        and the pressure value is fixed to $\beta pl_2^2=360$.
	}
	\label{fig5}
\end{figure}

\begin{figure*}
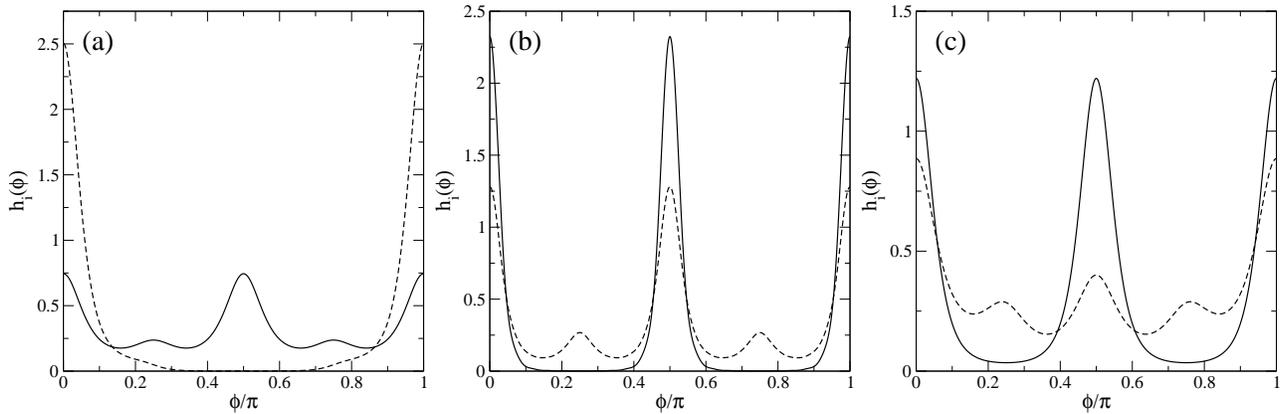

	\epsfig{file=fig5a.eps,width=2.2in}
	\epsfig{file=fig5b.eps,width=2.2in}
	\epsfig{file=fig5c.eps,width=2.2in}
	\caption{Orientational distribution functions for 
	squares, $h_1(\phi)$ (solid), and triangles, $h_2(\phi)$ (dashed), 
	corresponding to mixtures with $\kappa_l=1/\sqrt{2}$ ($\kappa_a=1$). 
	The molar fractions are fixed at (a) $x=0.1$, (b) $x=0.93$, and (c) $x=0.63665$. 
	The value of pressure is $\beta p l_2^2=360$.} 
	\label{fig6}
\end{figure*}

These phase diagrams resemble those recently obtained
from MC simulations of mixtures of hard disks and squares \cite{Escobedo2}, 
where the packing fraction for the I-hexatic transition
(counterpart of the present I-T transition) increases with the molar fraction 
of squares 
(disks). Both transition curves merge in a mosaic region where a micro-segregated phase 
with mixed hexatic and T symmetries become stable. This region is in turn
bounded above by a triangular-solid-T demixing which, at high enough pressures, becomes a solid-solid demixing between a triangular solid, rich in 
disks, and a square crystal, rich in squares. Here we do not 
consider nonuniform phases in the triangle-square mixture, so we cannot 
discard that, for high enough pressures, the demixed phases found are in fact
crystalline (instead of liquid crystalline).

We now study the orientational properties of the mixture with asymmetry $\kappa_l=1/\sqrt{2}$ 
($\kappa_a=1$). The fluid pressure was fixed to a value $\beta pl_2^2=360$, i.e. above
the I-N-T triple point (see panel (b) of Fig. \ref{fig4}). The total free-energy per particle 
was minimized with respect to all Fourier amplitudes $\{h_n^{(i)}\}$, and the equilibrium 
orientational distribution functions $\{h_i(\phi)\}$ and order parameters $Q_{2n}^{(i)}$ were
obtained as a function of mixture composition $x$. Results for the 
latter are shown in Fig. \ref{fig5}. Note that, for a wide range of compositions, 
the curves $Q_{2n}^{(i)}(x)$ represent order parameters of the unstable mixture due to the 
demixing transition shown in Fig. \ref{fig4}(b). However it is illustrative to look at
the behavior of the orientational ordering of triangles and squares as a function of 
composition for the whole interval $[0,1]$. Close to the one-component 
limits $x=0$ or $x=1$, the N order parameter of triangles $Q_2^{(2)}$, or the T order parameter 
of squares, $Q_4^{(1)}$, is highest, indicating strong N or T orientational ordering. 
In the neighborhood of $x=0$ squares follow the orientation of the more abundant triangular 
species by orienting their axes into a T configuration, but with rather low orientational ordering. 
However, as Fig. \ref{fig6}(a) indicates, the function $h_1(\phi)$ has (aside from the three main 
peaks located at $\{0,\pi/2,\pi\}$, typical of the T symmetry),
two additional small peaks at $\pi/4$ and $3\pi/4$; this is again an indication that square-triangle
interactions are behind the rising of orientational correlations with 
eightfold symmetry. This in turn affects the difference 
between the order parameters $Q_4^{(1)}$ and 
$Q_8^{(1)}$, which is anyway rather small as can be seen from Fig. \ref{fig5} (a).

For triangles, we see from Fig. \ref{fig5}(b) that $Q_2^{(2)}(x)$ decreases with $x$ and becomes 
zero for $x\ge x^*\simeq 0.65$, while $Q_4^{(2)}$ is always different from zero, exhibits a local 
minimum at $x^*$ and then increases monotonically. This means that for $x^*\leq x\leq 1$ triangles 
adopt the same T-orientational symmetry as squares. It is interesting to note that, for these
compositions, the values of the T ($Q_4^{(2)}$) and O ($Q_8^{(2)}$) order parameters of triangles 
are very similar, which again points to the existence of square-triangle correlations with 
eightfold symmetry. This feature can be directly seen in Fig. \ref{fig6} (b), where the functions 
$h_i(\phi)$ are plotted for a stable mixture with $x=0.93$. 
Note the strong T ordering of both, squares and triangles, indicated by the presence of sharp 
peaks at $\{0,\pi/2,\pi\}$, but also the presence of small satellite peaks in 
$h_2(\phi)$ at $\{\pi/4,3\pi/4\}$, a clear signature of the O-orientational correlations promoted 
by square-triangle interactions. 
Despite the presence of these correlations, we should bear in mind that triangles are clearly oriented
in a T-configuration, with the symmetry $h_2(\phi)=h_2(\phi+\pi/2)$. The exact O symmetry, 
$h_2(\phi)=h_2(\phi+\pi/4)$, is never observed for any mixture-asymmetry, pressure or composition values.

Fig. \ref{fig6}(c) shows the functions $h_i(\phi)$ for a mixture with a reference value of $x=0.63665$
(at which the curves $Q_2^{(2)}$ and $Q_4^{(2)}$ cross each other). 
Note that this mixture is unstable with respect to phase-separation. In this case squares clearly
orient in a T configuration, but triangles have a rather low uniaxial N ordering, with significant
secondary peaks at $\{\pi/4,\pi/2,3\pi/4\}$, again a signature of eightfold 
square-triangle orientational correlations. The conclusion can be extracted from all these results that,
despite the existence of square-triangle interactions, which promote the presence of small secondary 
peaks in $h_2(\phi)$ signalling O-type ordering, the effect is not sufficient to stabilize the O phase.

\section{The clustering effect}
\label{clustering}

\begin{figure}
	\epsfig{file=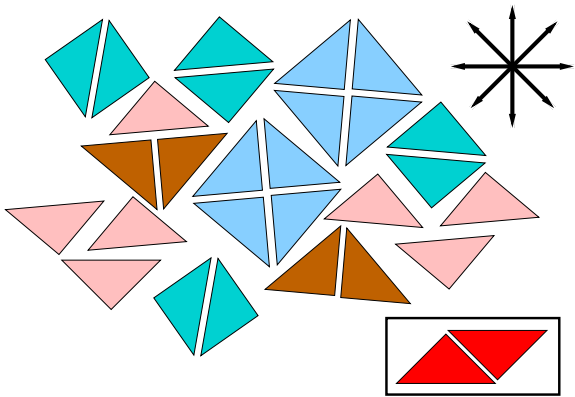,width=3.in}
	\caption{Sketch of three different clusters  
	and free monomers
	(shown in different colors) 
	formed by right-angled triangular monomers: (i) two different dimers are formed by 
	joining the largest or shortest sides of two triangular monomers,
	giving a square dimer or a (larger) right-angled triangle, respectively; (ii)
	tetramers are formed by joining four triangular monomers with
	their shortest sides in contact, producing a big square with side equal to the 
	largest side of the monomers. All clusters (species) are sketched in a 
	tetratic-like configuration, promoted by the square clusters, giving rise to a  
	global monomer orientation with an approximate eightfold symmetry (this is shown
	by eight arrows, which represent the eight equivalent directors). 
	A cluster with 
	rhomboidal symmetry, not taken into account in our toy model,
is shown enclosed in a rectangular box.}
	\label{fig7}
\end{figure}

As shown in our previous paper \cite{nosotros}, a DFT based on the second- or third-virial 
coefficients is not capable of accounting for the O liquid-crystal symmetry in a one-component
fluid of hard right triangles. MC simulations \cite{Dijkstra,nosotros}, however, indicate that 
these correlations are present. Clearly, any attempt to formulate a theory for the fluid of
hard right triangles should consider at least four-body particle correlations. Needless to say
this is an exceedingly complicated task. Because of these high-order correlations, we may expect 
particles to be prone to arrange into more or less short-lived clusters containing a few (but anomalously
high as compared to `normal' fluids) number of particles. 
In the case of right triangles it is not difficult to think of the geometries of the  
most stable particle arrangements (see below). It is also reasonable to expect that these `clusters'
may dominate the fluid structure and govern the bulk orientational properties of the fluid.
With this idea in mind, a step forward in our attempts to construct an alternative model for 
O correlations is based on considering these clusters as special entities, connecting to the idea 
self-assembly of particles (taken as monomer units). This idea leads to an
extreme model, where monomers form `superparticles', which in turn may orient in such a way as to
give rise to eightfold symmetry in the final monomer orientational distribution function.

Based on previous MC simulations \cite{Dijkstra,nosotros} and on additional MC simulations to be 
presented below, we have identified what can be regarded as `important' local particle configurations 
in the fluid at high packing fractions. A total of four such configurations
have been chosen because of their high probability along the MC chains generated in the simulations. 
A sketch of these four important configurations is shown in Fig. \ref{fig7}, where they are drawn 
in different colors. Hereafter these configurations will be called `clusters' and we now proceed
to defined their structure and shapes.

First we define `big-square' clusters. These appear close to the T-K transition and may be regarded as
tetramers, made out of four right triangles (monomers) with their short equally-sized sides 
(of length $l$) almost in contact, and with their right-angled vertexes also in close proximity. 
This configuration gives a big square with side equal to the triangle hypotenuse ($\sqrt{2}l$). 
Four-body correlations obviously induce the formation of these structures.
We also identify another type of cluster of square symmetry but with smaller size, obtained by 
joining two triangular monomers by their hypotenuse, creating a small square dimer of side $l$. 
Obviously the presence of these clusters in the fluid is very likely 
because this configuration guarantees the absolute minimum of the triangle-triangle 
exluded area (when the relative angle between particle axes is $\pi$), but note that tetramers do not
result by merging two of these clusters.
Next, if two triangular monomers are joined by their equally-sized sides with the 
right-angled vertexes in contact they form a large right-angled triangular 
cluster with the equally-sized lengths equal to $\sqrt{2}l$. Triangular clusters need to be
considered in the analysis since tetramers can be formed by merging two of these.
The last species to consider is, obviously, the free triangular monomer (the building block of 
all the larger clusters).

We should note that a rhomboidal dimer (see Fig. \ref{fig7})  
can be also formed by two triangles with their small sides almost in contact and their
right-angled and acute-angled vertexes in close proximity. Even though these clusters
are present in some cases (see later), they will be discarded from the model in order to make it
computationally manageable. Also, other possible clusters of different shapes or 
larger sizes will not be taken 
into account. Again we refer to Fig. \ref{fig7} for the definition of the four clusters and also to
Table \ref{tabla2} where their shapes, sizes and areas are summarized. 

In the following we present a simple extension of the SPT model for the quaternary mixture 
that results from a consideration of the four clusters defined above as distinct species. 
$l_i^{(k)}$ will denote
the length of the species, with $i$ indicating the geometry ($i=1$ for squares and $i=2$ for 
triangles), and $k$ the cluster size ($k=1$ for small and $k=2$ for big clusters). 
The excess free-energy per particle of the mixture can be obtained by
substituting the product $l_il_j$ by $l_i^{(k)}l_j^{(m)}$, and the particle areas $a_i$ by $a_i^{(k)}$,
into Eqn. (\ref{coefficients}), thus obtaining the generalized coefficients ${\cal K}_{ij,n}^{(2,km)}$ 
which are then used in (\ref{proposed}). Note that the sums over $ij$ in the latter 
equation, and also in the ideal free-energy (\ref{lo_ideal}), should run over four   
($ijkm$) and two ($ik$) indexes, respectively. Obviously we need to extend the same labelling to 
the molar fractions and to the orientational distribution functions: $x_i^{(k)}$ and $h_i^{(k)}(\phi)$,
respectively. The total packing fraction is then $\displaystyle\eta=\rho\sum_{ik}x_i^{(k)}a_i^{(k)}$, 
and the orientational order parameters are written as
\begin{eqnarray}
Q_{2n}^{(ik)}=\int_0^{2\pi}d\phi h_i^{(k)}(\phi)\cos(2n\phi)\hspace{0.8cm}(n=1,\cdots,4). 
\end{eqnarray}

\begin{table} 
	\begin{tabular}{|| c | c | c | c | c ||}
		\hline\hline
		Species $(ij)$ & \centering{11} & \centering{12} & \centering{21} & 22\\
		\hline
		Geometry & 
		\ \includegraphics[width=0.3in]{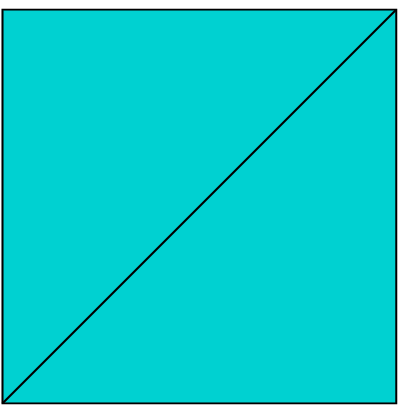} \
		& \ \includegraphics[width=0.4in]{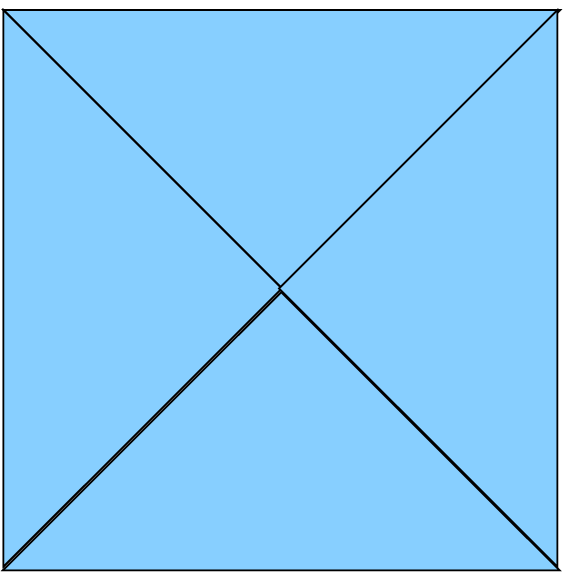} \
		&  \ \includegraphics[height=0.2in]{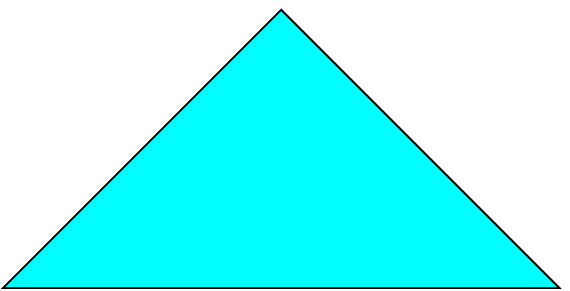} \
		&  \  \includegraphics[height=0.28in]{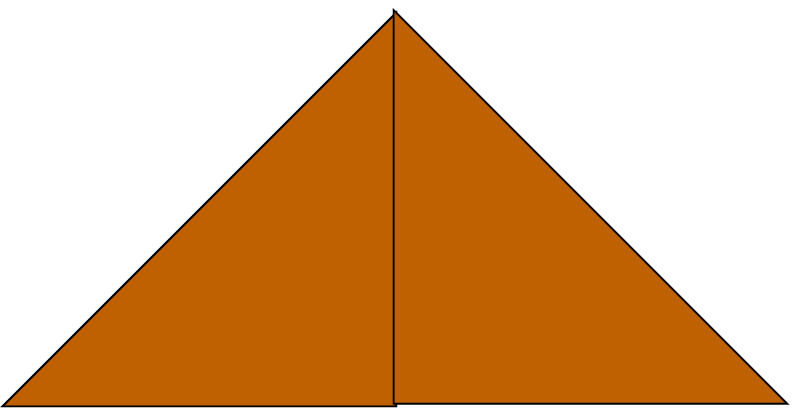} \  \\
		\hline
		$l_i^{(j)}$ & $l$ & $\sqrt{2}l$ & 
		$l$ & $\sqrt{2}l$ \\
		\hline
		$a_i^{(j)}$ & $l^2$ & $2l^2$ & 
		$l^2/2$ & $l^2$\\
		\hline\hline
	\end{tabular}
	\caption{Labelling of the different species with their shapes,
	characteristic lengths, $l_i^{(j)}$, and areas, $a_i^{(j)}$. The unit is the 
	monomer length $l$.}
	\label{tabla2}
\end{table}

\begin{figure}
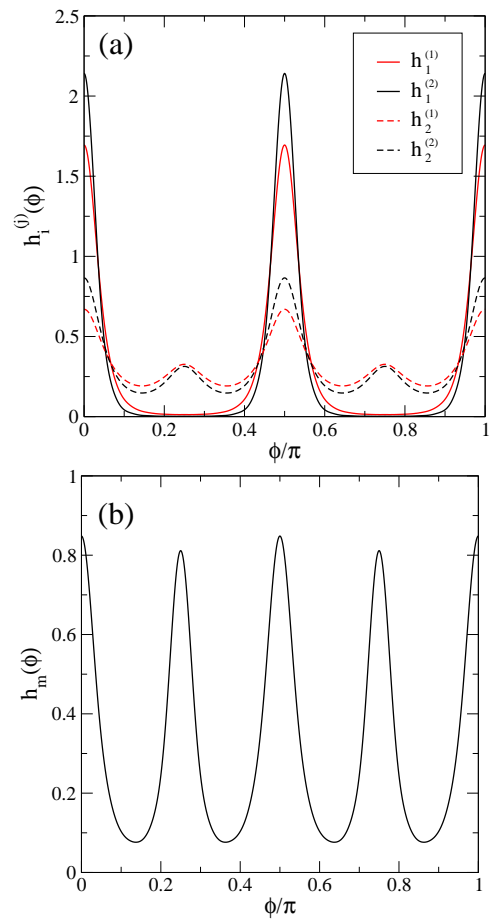

	\epsfig{file=fig7a.eps,width=2.5in}
	\epsfig{file=fig7b.eps,width=2.43in}
	\caption{(a) Equilibrium orientational distribution functions 
	$h_i^{(j)}(\phi)$ ($i,j=1,2$) of different 
	species (clusters) in a quaternary mixture with molar fractions 
	$x_1^{(1)}=0.4$, $x_1^{(2)}=0.15$ (small and big squares, respectively), $x_2^{(1)}=0.35$, 
	and $x_2^{(2)}=0.1$ (small and big triangles, respectively). The results were obtained 
	using the SPT. The scaled pressure is fixed to $p^*=220$. (b) 
	Orientational distribution function for the monomer axis orientation, 
	$h_{\rm m}(\phi)$, as obtained from Eqn. (\ref{monomer}). The approximate eightfold 
	symmetry of $h_{\rm m}(\phi)$ is clearly seen.}
	\label{fig8}
\end{figure}

\begin{figure}
	\epsfig{file=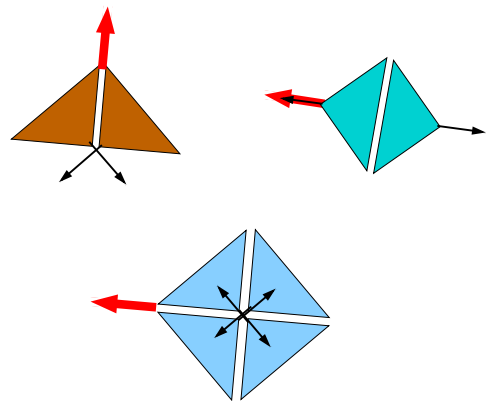,width=2.5in}
	\caption{Sketch of triangular and square dimers and tetramers, with the main monomer 
	(black thin arrows) and cluster (red thick arrows) axes shown.}
	\label{new_figure}
\end{figure}

Now we specify how the orientational ordering of monomers is calculated from that of clusters.
Cluster numbers in the mixture are given by $n_i^{(k)}=x_i^{(k)}N_{\rm c}$ ($i,k=1,2$), 
with $N_{\rm c}$ the total number of clusters. The total 
number of monomers distributed among all different clusters can be calculated as 
$N_{\rm m}=\left(2x_1^{(1)}+4x_1^{(2)}+x_2^{(1)}+2x_2^{(2)}\right)N_{\rm c}$. 
First let us consider the case of square dimers. The main axes of the triangular monomers 
point parallel and antiparallel to one of the square axis, 
i.e. the one perpendicular to the square diagonal coinciding with the monomer hypotenuse 
(see Fig. \ref{new_figure}). 
Note that if we select the square axis to be parallel to the other diagonal these angles 
are $\pm \pi/2$. However, 
as the squares only have I or T liquid-crystal phases, which are orientationally
invariant with respect to $\pi/2$ rotations, the above definitions are identical, i.e. 
they do not affect the final result for the orientational ordering of monomers.
Thus the contribution of the $n_1^{(1)}$ square dimers to the global orientational 
ordering of monomers is given by the function 
\begin{eqnarray}
	&&h_{\rm m}^{(11)}(\phi)=\frac{n_1^{(1)}}{N_{\rm m}}\left(h_1^{(1)}(\phi)+
	h_1^{(1)}(\phi+\pi)\right)\nonumber\\
	&&=\frac{x_1^{(1)}\left(h_1^{(1)}(\phi)+h_1^{(1)}(\phi+\pi)\right)}
	{2x_1^{(1)}+4x_1^{(2)}+x_2^{(1)}+2x_2^{(2)}}.
\end{eqnarray}
For square tetramers, we can see, as Fig. \ref{new_figure} shows, that the axes 
of the triangular monomers are at
angles $\{\pi/4,-\pi/4,3\pi/4,-3\pi/4\}$ with respect to one of the square diagonals. Thus 
the contribution of the $n_1^{(2)}$ square tetramers is
\begin{eqnarray}
	h_{\rm m}^{(12)}(\phi)=\frac{x_1^{(2)}
	\sum_{k=\pm 1}\left(h_1^{(2)}(\phi+k\pi/4)+
	h_1^{(2)}(\phi+3k\pi/4)\right)}
	{2x_1^{(1)}+4x_1^{(2)}+x_2^{(1)}+2x_2^{(2)}}.\nonumber\\
\end{eqnarray}
The $n_2^{(1)}$ free triangular monomers give a contribution of
\begin{eqnarray}
	h_{\rm m}^{(21)}(\phi)=\frac{x_2^{(1)}h_2^{(1)}(\phi)}
	{2x_1^{(1)}+4x_1^{(2)}+x_2^{(1)}+2x_2^{(2)}}.
\end{eqnarray}
Finally, for triangular dimers, it is easy to see that the two monomer axes point at angles 
$\{3\pi/4,-3\pi/4\}$ with respect to the main axis of the dimer  
(see Fig. \ref{new_figure}). 
The contribution of the $n_2^{(1)}$ 
triangular dimers is then
\begin{eqnarray}
	h_{\rm m}^{(22)}(\phi)=\frac{x_2^{(2)}\left(h_2^{(2)}(\phi-3\pi/4)+
	h_2^{(2)}(\phi+3\pi/4)\right)}{2x_1^{(1)}+4x_1^{(2)}+x_2^{(1)}+2x_2^{(2)}}.
\end{eqnarray}
The total orientational distribution function of monomers is just the sum of the different contributions 
obtained above, 
\begin{eqnarray}
	h_{\rm m}(\phi)=\sum_{i,j=1,2}h_{\rm m}^{(ij)}(\phi),
	\label{monomer}
\end{eqnarray}
and from this we can calculate the order parameters of monomers as 
\begin{eqnarray}
	Q_{2n}^{(\rm m)}=\int_0^{2\pi}d\phi h_{\rm m}(\phi)\cos(2n\phi).
	\label{q_monomer}
\end{eqnarray}
We have performed a minimization of the total free-energy per particle with respect 
to all the Fourier amplitudes $\{h_n^{(ik)}\}$ (note the labelling extension $i,k=1,2$) of 
the quaternary mixture. Possible demixing scenarios were not searched for because we are 
only interested in the effect of clustering on the orientational ordering of monomers. 
Fig. \ref{fig8}(a) shows the equilibrium orientational distribution 
functions $\{h_i^{(j)}(\phi)\}$, for a scaled 
pressure fixed to $p^*\equiv\beta p \left(l_2^{(1)}\right)^2=220$, 
and a set of molar fractions with values $x_1^{(1)}=0.4$, $x_1^{(2)}=0.15$, $x_2^{(1)}=0.35$ and 
$x_2^{(2)}=0.1$, fulfilling the equality $x_1^{(1)}+x_2^{(2)}=x_1^{(2)}+x_2^{(1)}=0.5$. Clearly the system
exhibits T ordering in all the species, with square clusters being more ordered and 
with the presence of secondary peaks (located at $\phi=\pi/4$ and $\phi=3\pi/4$) in the distribution 
functions of triangular clusters. As explained above, this is due to square-triangle interactions. 
The monomer distribution function $h_{\rm m}(\phi)$, calculated from Eqn. (\ref{monomer}), is shown in 
panel (b). It has a quasi-eightfold symmetry, with peaks located at $k\pi/4$ ($k=1,\cdots,4$). 
Note however that the perfect symmetry $h_{\rm m}(\phi)=h_{\rm m}(\phi+\pi/4)$ is not exactly 
fulfilled: small differences in the height of the peaks are clearly visible. To put this result in
perspective we remark that, in a MC simulation, small differences like these would naturally be
attributed to the effect of limited angular sampling in the histogram of $h(\phi)$. In any case the 
function $h_{\rm m}(\phi)$ plotted in Fig. \ref{fig8}(b) shows a high O-type ordering, which is a
direct consequence of the strong particle clustering.

\begin{figure*}
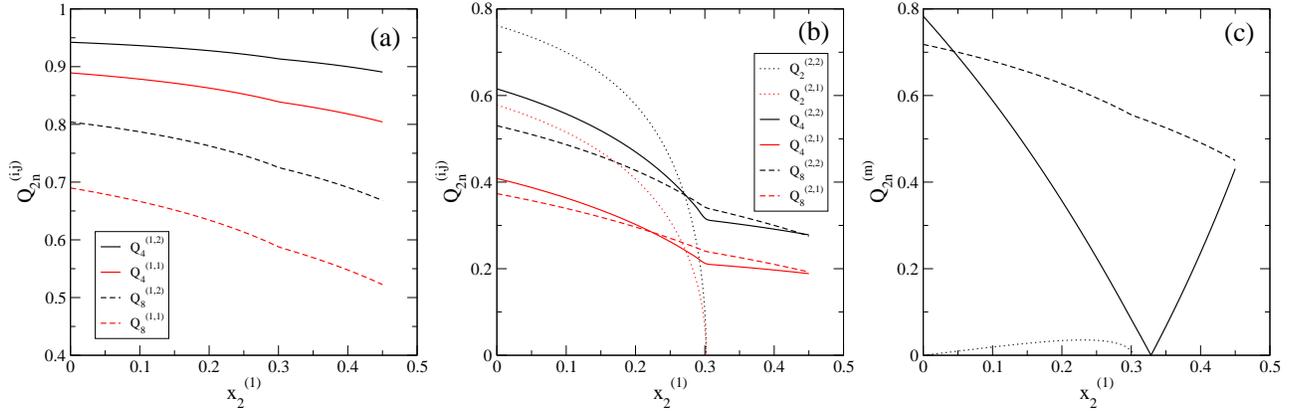

	\epsfig{file=fig9a.eps,width=2.2in}
	\epsfig{file=fig9b.eps,width=2.2in}
	\epsfig{file=fig9c.eps,width=2.2in}
	\caption{N ($n=1$), T ($n=2$) and O ($n=4$) 
	order parameters, $Q_{2n}^{(i,j)}$, of (a) square-like species, $i=1$ and $j=1,2$, 
	and (b) triangular species, $i=2$ and $j=1,2$, as a function of the free monomer composition,        $x_2^{(1)}\in[0,0.45]$. The order parameters were obtained from the SPT-minimization of 
	a quaternary mixture, at fixed scaled pressure $p^*=220$, and with 
	species compositions $x_1^{(1)}=x_2^{(1)}+0.05$, $x_1^{(2)}=0.5-x_2^{(1)}$, and
        $x_2^{(2)}=0.45-x_2^{(1)}$, which fulfill the constraint 
	$x_1^{(1)}+x_2^{(2)}=x_1^{(2)}+x_2^{(1)}=0.5$. 
	(c) Monomer order parameters, $Q_{2n}^{(\rm m)}$,
	as calculated from Eqn. (\ref{q_monomer}), 
	using the function $h_{\rm m}(\phi)$, 
	for $n=1$ (dotted curve), $n=2$ (solid curve) and $n=4$ (dashed curve). 
	Note that, in the case $n=2$, it is the absolute value of the T order parameter, 
	$|Q^{(\rm m)}_4|$, that is shown (this is due to the change of sign 
	at $x_2^{(1)}\approx 0.33$).} 
	\label{fig9}
\end{figure*}

Next, order parameters $Q_{2n}^{(ij)}$ obtained using Eqn. (\ref{q_monomer}) are shown in 
Fig. \ref{fig9} as a function of monomer composition $x_2^{(1)}\in[0,0.45]$, following a path in
molar fractions $x_1^{(1)}=x_2^{(1)}+0.05$, $x_1^{(2)}=0.5-x_2^{(1)}$ and $x_2^{(2)}=0.45-x_2^{(1)}$ 
(these values fulfill the constraints $x_1^{(1)}+x_2^{(2)}=x_1^{(2)}+x_2^{(1)}=0.5$:
the sum of compositions of big triangles and small  squares is equal to the sum of compositions 
of the other two species). Panel (a) shows that square tetramers have the largest T ordering, 
as compared to that of square dimers. This order decreases with $x_2^{(1)}$ as a consequence of the 
fact that $x_1^{(2)}$ (the fraction of big squares) also decreases along the selected path. In turn, triangular dimers and monomers 
exhibit N ordering up to $x_2^{(1)}\simeq 0.3$, beyond which it vanishes due to the fact that 
$x_2^{(2)}$ (the fraction of big triangles) 
decreases with $x_2^{(1)}$ along the same path, see panel (b). From this value, triangular species 
follow the T ordering of the square species. It is interesting to note that the O order parameter 
of the triangular species, $Q_8^{(2j)}$ ($j=1,2$), becomes larger than the T order parameter,
$Q_4^{(2j)}$, indicating the presence of satellite peaks in $h_2^{(j)}(\phi)$ 
at $\pi/4$ and $3\pi/4$. Panel (c) shows the monomer order parameters $Q_{2n}^{(\rm m)}$ as 
a function of $x_2^{(1)}$, calculated from Eqn. (\ref{q_monomer}). The N ordering of monomers is 
relatively low, something that can be seen from the negligible value of $Q_2^{(\rm m)}$ 
which becomes zero beyond $x_2^{(1)}=0.3$. 
We also see that, close to $x_2^{(1)*}\simeq 0.33$, the T order parameter, 
$Q_4^{(\rm m)}$, becomes zero,
while the O ordering, measured through $Q_8^{(\rm m)}$, is relatively high in the neighborhood of $x_2^{(1)*}$. 
Therefore there exists an interval in $x_2^{(1)}$ around $x_2^{(1)*}$ where the orientational distribution 
function of monomers $h_{\rm m}(\phi)$ is similar to that shown in Fig. 
\ref{fig8} (b), i.e. it shows
a quasi-eightfold symmetry. Note that, for $x_2^{(1)}<x_2^{(1)*}$, the T director of square clusters  
coincides with that of the preferential alignment of square dimers, while for $x_2^{(1)}>x_2^{(1)*}$ 
it changes to that of square tetramers (rotated by $\pi/4$ with respect to the former). This is the reason
why the order parameter $Q_4^{(\rm m)}$ exhibits a change in sign at $x_2^{(1)*}$. 
Again we can conclude from these 
results that the O ordering is highly enhanced by the presence of particle clustering: when 
monomers are mainly distributed into square dimers and tetramers, there exists an interval 
in the composition of free monomers for which the monomer distribution function exhibits quasi-eightfold 
symmetry. 

\begin{figure*}
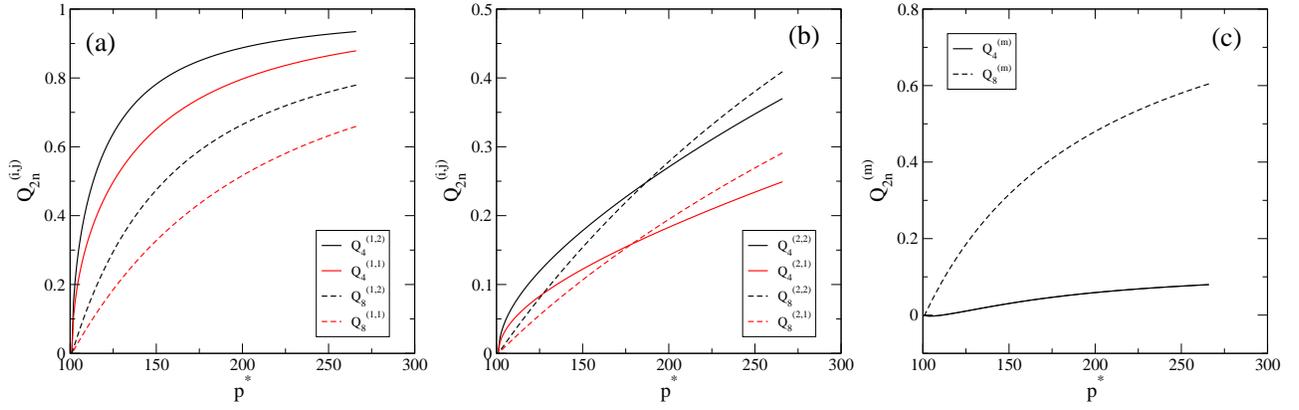

	\epsfig{file=fig10a.eps,width=2.2in}
	\epsfig{file=fig10b.eps,width=2.2in}
	\epsfig{file=fig10c.eps,width=2.2in}
	\caption{Order parameters $Q_{2n}^{(i,j)}$ as a function of reduced pressure 
	$p^*$ of (a) square and (b) triangular species for a quaternary mixture with  
	fixed compositions $x_1^{(1)}=0.4$, $x_1^{(2)}=0.15$, $x_2^{(1)}=0.35$,
	and $x_2^{(2)}=0.1$. (c) The resulting T ($n=2$, solid curve) and O ($n=4$, 
	dashed curve) monomer order parameters, $Q_{2n}^{(\rm m)}$.}
	\label{fig10}
\end{figure*}

Fig. \ref{fig10} shows the evolution of the order parameters of all species 
[panels (a) and (b]) and of the monomers [panel (c)] as a function of reduced pressure 
$p^*$ for the same fixed set of compositions as in Fig. \ref{fig8}: 
$x_1^{(1)}=0.4$, $x_1^{(2)}=0.15$, $x_2^{(1)}=0.35$, and $x_2^{(2)}=0.1$, where all
species exhibit T ordering. The quasi-O ordering, measured by $Q_8^{(\rm m)}$,
increases from zero at the same pressure value where square dimers and tetramers 
exhibit a second-order I-T transition at 
$\beta p l_2^2\approx 100$. 
For higher pressures the O 
order parameter of monomers, $Q_8^{(\rm m)}$, is significantly larger than the T order parameter $Q_4^{(\rm m)}$.

\begin{figure}
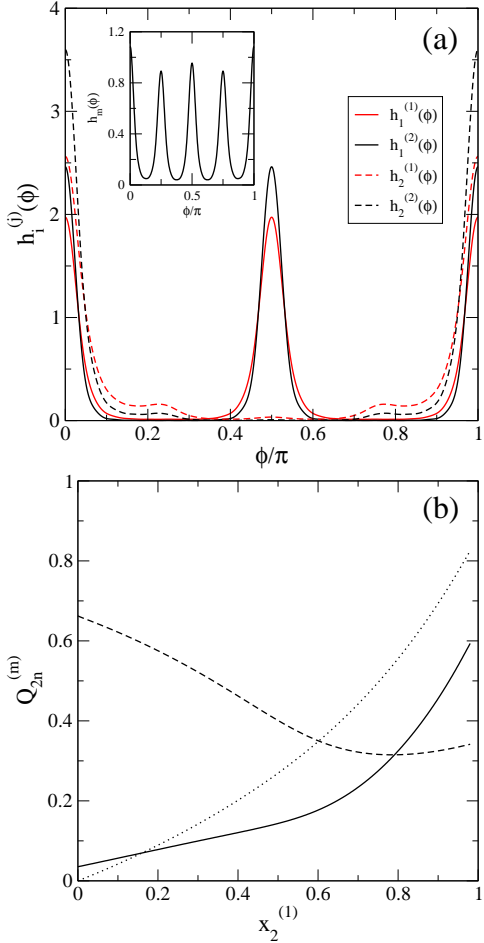

	\epsfig{file=fig11a.eps,width=2.5in}
	\epsfig{file=fig11b.eps,width=2.5in}
	\caption{(a) Angular distribution functions of clusters 
	$h_i^{(j)}(\phi)$ in the 
	quaternary mixture with compositions $x_1^{(1)}=0.45$, $x_1^{(2)}=0.01$, 
	$x_2^{(1)}=0.1$ and $x_2^{(2)}=0.44$, and reduced pressure $p^*=300$. 
	The inset corresponds to the 
	angular distribution function of monomers, $h_{\rm m}(\phi)$. 
	(b) Order parameters of monomers, 
	$Q_{2n}^{(\rm m)}$, for $n=1$ 
	(dotted curve), $n=2$ (solid curve) and 
	$n=4$ (dashed curve), 
	as a function of the free monomer composition $x_2^{(1)}\in[0,0.98]$, 
	following a path $x_1^{(1)}=0.5-x_2^{(1)}/2$, $x_1^{(2)}=0.01$, 
	and $x_2^{(2)}=0.49-x_2^{(1)}/2$. The pressure is fixed to the same 
	value as in (a).}
	\label{fig11}
\end{figure}

The angular distribution function of monomers with quasi eightfold-symmetry can also be obtained 
for a ternary mixture of free monomers and dimers of triangular or square symmetry, i.e. for 
vanishingly small tetramer composition. This is shown in Fig. \ref{fig11}(a), where all 
cluster orientational distribution functions, $h_i^{(j)}(\phi)$, are shown
for a quaternary mixture with $x_1^{(1)}=0.45$, 
$x_1^{(2)}=0.01$, $x_2^{(1)}=0.1$ and $x_2^{(2)}=0.44$ and reduced pressure                        
$p^*=300$. For comparison, the monomer function $h_{\rm m}(\phi)$ is also shown (see inset). 
This time free monomers and triangular dimers clearly orient in a N-like 
configuration. As pointed out before, the axes of monomers in triangular dimers are oriented 
with respect to the dimer axis with relative 
angles of $\pm 3\pi/4$, while monomers in square dimers have their axes 
parallel or anti-parallel to the dimer axis. Taking into account these relative orientations,
and the fractions of the different clusters ($0.45$ and $0.44$ for triangular and
square dimers, respectively; and the rather small values of $0.1$ and $0.01$
for free monomers and tetramers, respectively), the quasi-eightfold symmetry of $h_{\rm m}(\phi)$
explains itself. We expect that the inclusion of the fourth virial coefficient in the theory would cause the orientational distribution function of triangles to exhibit strong O correlations, since configurations where
two or four triangles form triangular or square dimers, and also square 
tetramers, might be entropically favored. 

The above set of values $\{x_i^{(j)}\}$ is just a particular case of the path 
obtained by varying the free-monomer composition $x_2^{(1)}$ inside the interval 
$[0,0.98]$, together with the constraints $x_1^{(1)}=0.5-x_2^{(1)}/2$, $x_1^{(2)}=0.01$, 
$x_2^{(2)}=0.49-x_2^{(1)}/2$ (keeping fixed the small tetramer composition). The 
evolution of the order parameters of monomers, $Q_n^{(\rm m)}$, with respect to 
$x_2^{(1)}$ along this path and for the same pressure $p^*=300$ is shown in 
Fig. \ref{fig11}(b). A wide region exists, close to the $x_2^{(1)}=0$ axis, where the 
O order parameter $Q_8^{(\rm m)}$ is highest, which corresponds to strong 
eightfold orientational correlations between monomers. In the opposite region, close to 
the $x_2^{(1)}=1$ axis, monomers are oriented in an N-like configuration  
($Q_2^{(\rm m)}$ begin the highest parameter). Obviously this is a direct consequence of the small 
population of triangular and square clusters with respect to free monomers.

\section{Monte Carlo simulations}
\label{MC}

To understand the relevance of clusters in the configurations of hard right 
triangles and give some support to the assumptions underlying the 
models presented in the previous sections, we have performed NVT-MC simulations of a system of 576 particles in a square box using periodic boundary conditions, using $2\times 10^5$ MC steps for equilibration and
$3\times 10^5$ MC steps for averaging. Different expansion and
compression runs were performed, starting from different initial 
configurations, to explore different liquid-crystalline phases. For
more details on the simulations we refer to our previous work 
\cite{nosotros}.

As shown in Ref. \onlinecite{nosotros}, the high-density fluid of hard right
triangles seems to be very prone to staying in specific configurations which
can be controlled by an adequate choice of symmetry for the initial 
configuration. This would mean that there are dense regions in phase space 
from which it would be difficult to escape, probably due to unlikely
local rearrangements of particle orientations. Thermodynamically we could think
of these configurations as corresponding to metastable phases. A reasonable
procedure to identify the true stable phase at each density would be to 
perform free-energy calculations using thermodynamic integration or
applying the coupling method \cite{Frenkel1}. In this section, however, we
are not interested in thermodynamic stability (which would require extensive
simulations), but rather we use this feature of the hard right-triangle
fluid to probe for particle clustering in fluids of different bulk symmetries. 
Also, since we are using the MC technique, we are not probing the stability
in time of clusters as separate entities, but simply the occurrence of
a set of particular configurations and their relative importance along the
MC chains.

While the definition of clusters in a one-component fluid of hard particles
may be clearly specified, the criteria used in a simulation to associate a 
particular local configuration of particles to a given cluster type is 
somewhat arbitrary. Here we have focused on the clusters defined in Section 
\ref{clustering}
used to explore the consequences of the mixture model since, as explained
previously, these are the most natural configurations of the system.
We advance that indeed these configurations are very frequent, depending on
the total fluid density.

Again we refer to Fig. \ref{fig7} and Table \ref{tabla2} for the definition
of three types of clusters: square tetramers, square dimers, an triangular
dimers. In the simulations we have also focused on rhomboidal dimers 
(defined in Section \ref{clustering}), 
since they may be present at not too high densities. 
To define a dimer with a particular symmetry
(either square, triangular or rhomboidal), we calculate the distance
between the barycenters of two neighbouring triangles as well as their 
relative angle. Each perfect dimer (particles in contact with the correct
relative orientation) has specific values for these two variables
(relative distance and relative angle). We
define a dimer of a given type whenever these variables depart by less than
15\% from their ideal values. This criterion is totally arbitrary and, in 
fact, may lead to a situation where some pairs of particles are not 
considered to form a dimer, whereas visually one would clearly ascribe the
pair to be a dimer. Also, a given pair of neighboring particles may be
considered as a dimer discontinuously along the MC chain. Finally, the 
fact that the same tolerance is used to define all types of dimers may
introduce a bias in the relative fraction of dimers. Again, the analysis
is qualitative and not aimed at extracting definite numbers on quantities
that are otherwise ill-defined.

Finally, square tetramers are defined as the association of two triangular
dimers by applying a more relaxed criterion on distance and angle (20\%)
with respect to the ideal values for a square tetramer with all four 
particles in contact. This is because the positions and angles of the two 
particles of a triangular dimer will already depart from the ideal values.

\begin{figure}
	\epsfig{file=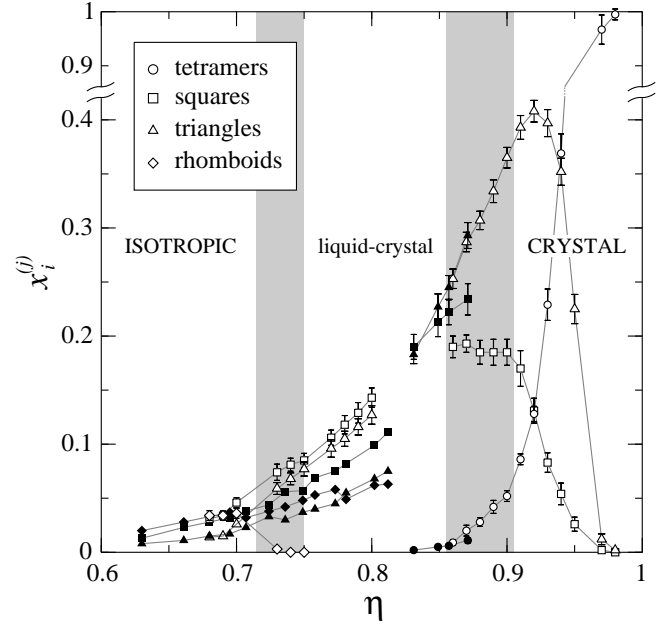,width=3.5in}
	\caption{Fraction of clusters $x_i^{(j)}$ obtained from the MC simulations
	as a function of packing fraction $\eta$. The nature of the cluster is indicated in the keybox. Filled symbols correspond to 
	compression runs, while open symbols are from expansion runs. The
	vertical shaded regions are approximate two-phase regions derived
	from data in Refs. \onlinecite{Dijkstra,nosotros}. Error bars are not shown when their sizes 
	are similar to the symbol sizes.}
	\label{fig12}
\end{figure}

Fig. \ref{fig12} shows the fractions of the four species of clusters as a 
function of packing fraction. To calculate the fraction of a given species,
the average number of clusters of that type along the simulation is
divided by the average number of clusters of all species, including "isolated"
triangles (monomers). Several expansion and compression runs were performed,
following the results presented in Ref. \onlinecite{nosotros}. In Fig. \ref{fig12}
the different clusters are indicated by different symbols, 
as shown in the keybox. Filled symbols correspond to
        compression runs, while open symbols are from expansion runs.
In the isotropic phase, our cluster criterion identifies a small fraction 
of the two types of dimers and rhomboids. They correspond to local
arrangements that do not persist along the MC chain. As density is increased
(compression run, filled symbols) cluster fractions also increase. In the
liquid-crystal region an order-parameter analysis (see Ref. \onlinecite{nosotros})
identifies this phase as tetratic or octatic, since the $Q_4$ and $Q_8$
adopt comparable and relatively high values. The strong 
O ordering is coming mainly from square dimers, which also force 
neighboring monomers to adopt orientations that foster this type of
ordering. Note that rhombic clusters are also present, but no tetramers
can be identified.

An expansion run from a perfect crystal of tetramers (perfect tetratic
crystal) at $\eta=0.98$ is also shown in Fig. \ref{fig12}. Initially only
tetramers are present, but as density is decreased these clusters break 
into triangular dimers. At the end of the crystal phase the latter
clearly dominate, but at the same time a I fraction
of square dimers is created. These are probably formed by "free" monomers 
that have been detached from neighboring tetramers. In essence the existence
of these clusters and the evolution of their fractions with density are
perfectly compatible with a crystal phase showing thermal fluctuations.
As expected, the fraction of rhombic clusters is essentially negligible
as particles would have to rotate locally by $90^{\circ}$, which is 
very difficult at high packing fraction.

Melting of the crystal is associated with a change in the variation of fractions with 
density. Tetramers have disappeared while the fractions of square
and triangular clusters tend to be similar. The fluid becomes ordered in
a T phase, dominated by these two types of clusters.
As density is further decreased this phase changes to the isotropic phase.
Note that in the T phase no rhombic clusters are excited. However, the
fraction of these clusters increases suddenly, and the equilibrium
value of this fraction in the isotropic phase is 
reproduced by the expansion run from the T as soon as the phase transition
is crossed.

A final compression run was performed from the T phase. As expected, 
the fluid cannot crystallize and the fraction of square clusters does
not match the one obtained from the expansion run, even though 
triangular clusters do have almost identical fractions.

In summary, the MC simulations show that the particle clusters introduced
in Sections \ref{mixtures} and \ref{clustering} do occur in the fluid
of hard right triangles, in high proportion and in varying degrees 
according to the global orientational order of the system. 
 Therefore, a model based
on the equilibrium statistics of these clusters as separate entities
may be a fruitful way to understand the essential orientational
properties of the hard right triangle fluid.

\section{Conclusions}
\label{conclusions}

In this paper we have addressed the origin of the liquid-crystal phase
of hard right triangles. Compression Monte Carlo simulations indicate 
the existence of an exotic liquid-crystalline phase exhibiting tetratic order 
and strong octatic correlations, which the
standard and extended SPT versions of DFT are unable to describe. 
As a step forward, and in view of the apparently important clustering 
tendencies of right triangles into square and triangular clusters, 
we have implemented an SPT (second-virial) approach to analize
the phase behavior of four different binary mixtures of right triangles and 
squares, and calculated their respective phase diagrams. The length-asymmetry 
of the species are $\kappa_{\rm l}=1/2$, $1/\sqrt{2}$, 1 and $\sqrt{2}$, 
where the second case corresponds to species having equal particle areas 
($\kappa_{\rm l}=1/\sqrt{2}$). 
All mixtures (including the equal-area one) exhibit strong I-T and N-T 
demixing, a region of I-N first order transition, and the presence of a 
I-N-T triple point. 
The demixing scenarios directly follow from the
different liquid-crystal symmetries exhibited by the one-component fluids 
of hard triangles and hard squares, i.e. N and T symmetries, respectively. 
Triangles in the mixture can adopt a T ordering which follows the natural 
ordering of the more populated square species, but with an orientational 
distribution function, $h_2(\phi)$, exhibiting two relatively small 
satellite peaks located at $\{\pi/4,3\pi/4\}$, which points to the 
importance of square-triangle interactions in the existence of orientational 
correlations with eightfold symmetry. However, the distribution function 
has a clear T character (with T and O order parameters having similar values). 

We also provided some results that explain the evidence that,
under certain conditions, particle clustering can give rise to situations
where the O order parameter, $Q_8$, is much higher that the N, $Q_2$, or T, 
$Q_4$ order parameters. Our results are based on the implementation of
a toy model consisting of a quaternary mixture where species are triangular 
monomers, triangular and square dimers, and finally tetramers, all assumed
to form in the real one-component fluid by monomer self-assembling. 
Evidence form MC simulation for the prevalence of these clusters in the 
equilibrium configurations of this fluid was presented in Section \ref{MC}.
We then used the SPT approach to estimate the free-energy of the mixture and, 
via minimization, calculated the equilibrium orientational distribution functions of clusters. From them the corresponding function for monomers can be derived.
The question of the possible demixing scenarios, which do certainly exist
in these mixtures, was not addressed. The focus was put on the monomer
distribution function $h_{\rm m}(\phi)$ which, for certain sets of cluster 
compositions, exhibits quasi eightfold symmetry, with four peaks of similar, 
although not identical, height in the interval $[0,\pi]$. This demonstrates
that the elusive eightfold ordering seen in the simulations, but not in the
standard two- and three-body versions of DFT, can originate from the 
prevalence of particle clustering and its effect on the global
orientational properties of the fluid. Further studies focusing on the 
dynamics of these processes may reveal whether the clustering idea is
just a convenient artifact to partition configurational space or a real
situation with clearly separated time scales associated to cluster kinetics,
internal cluster dynamics and the fluid dynamics of clusters as separate 
entities. In any case, our results indicate that it is reasonable to appeal to 
the clustering effect to explain why the O phase, observed in simulations, 
cannot be stabilized by the usual implementations of DFT. 

Even if the idea of the fluid as a collection of clusters turns out to be useful, 
the present assumption that cluster composition can be fixed in advance 
should be improved by considering these compositions as an output of the model.
Consistent with the theory for the kinetics of clustering,
monomer aggregation and evaporation, and cluster formation and fragmentation,
may be described by a set of chemical reactions, with certain reaction 
constant ratios at equilibrium. These ratios can be obtained from the difference between the chemical potentials of the species involved in the reactions. 
The more realistic model would consider the fluid as a 
polydisperse mixture of cluster or superparticles of different sizes and shapes,
each cluster having a particular association energy as in models of 
associated fluids. We believe these ideas deserve some exploration in the future.

\acknowledgements

Financial support under grant FIS2017-86007-C3-1-P
from Ministerio de Econom\'{\i}a, Industria y Competitividad (MINECO) of Spain,
and PGC2018-096606-B-I00 from Agencia Estatal de Investigaci\'on-Ministerio de Ciencia 
e Innovaci\'on of Spain,
is acknowledged.

\end{document}